\documentclass[aps,pre,reprint,showpacs,groupedaddress]{revtex4-2}

\usepackage{amsmath}
\usepackage{amssymb}
\usepackage{hyperref}
\usepackage{graphicx}
\usepackage{cleveref}
\hypersetup{
    colorlinks,
    citecolor=blue,
    linkcolor=blue,     urlcolor=blue
}

\newcommand{\vecthree}[3]{\left( \begin{array}{c} #1\\#2\\#3\end{array} \right)}
\newcommand{\vecfour}[4]{\left( \begin{array}{c} #1\\#2\\#3\\#4\end{array} \right)}
\renewcommand{\sp}[1]{\hat{\mathcal{#1}}}
\renewcommand{\vec}[1]{\boldsymbol{#1}}

\begin{document}


\title{Absorption of charged particles in Perfectly-Matched-Layers
by optimal damping of the deposited current}

\author{Remi Lehe$^a$}
\email[]{rlehe@lbl.gov}
\author{Aurore Blelly$^a$}
\thanks{Now at IRFU, CEA, Universit\'e Paris-Saclay, France}
\author{Lorenzo Giacomel$^b$}
\author{Revathi Jambunathan$^a$}
\author{Jean-Luc Vay$^a$}
\affiliation{$^a$ Lawrence Berkeley National Laboratory, Berkeley, CA 94720, USA\\
$^b$ CERN - 1211 Geneva 23 - Switzerland}

\date{\today}

\begin{abstract}
Perfectly-Matched Layers (PML) are widely used in Particle-In-Cell simulations,
in order to absorb electromagnetic waves that propagate out of
the simulation domain. However, when charged particles cross the interface
between the simulation domain and the PMLs, a number of numerical artifacts
can arise. In order to mitigate these artifacts, we introduce a new PML algorithm
whereby the current deposited by the macroparticles in the PML is damped
by an analytically-derived, optimal coefficient. The benefits of this
new algorithm is illustrated in practical simulations. In particular, it is shown that this
new algorithm is well-suited for particles exiting the box in near-normal incidence -- in the sense that the fields behave as if the exiting particle was propagating in an infinite vacuum.
\end{abstract}

\maketitle

\section*{Introduction}

Electromagnetic Particle-In-Cell (PIC) simulations \cite{Birdsall2004,Hockney1988}
are widely used to study the physics of beams and plasmas in various
contexts, including astrophysical systems, particle accelerators,
and microwave devices. In those different cases, the simulation domain
represents only a finite portion of space. Therefore, electromagnetic
waves and charged particles can potentially exit
that domain during the course of the simulation.
If no specific treatment is applied at the domain boundary, this can
lead to unphysical effects, such as electromagnetic waves reflecting back into
the interior of the domain, and charged particles leaving spurious fields
at the boundary.
Oftentimes, these effects can invalidate the results of the simulation.
Hence it is necessary to apply algorithms that \emph{``remove''}
(or \emph{``absorb''}) out-going electromagnetic waves and charged particles,
in a way that reproduces the expected physical behavior.

For electromagnetic waves, one such algorithm is the Perfectly-Matched Layers
(PMLs) algorithm \cite{Berenger_JCP_1994}. PMLs are auxiliary
computational cells where the field update equations are modified so as to damp
out-going electromagnetic waves without undesired reflection. This technique has indeed been shown to efficiently
absorb waves over a broad range of incidence angles and frequencies, and is
therefore very commonly used. However, in the original formulation of the PMLs,
the presence of charged particles was not considered. In practice, a number of
numerical issues can occur when charged particles reach the PML region, which can
then undesirably affect the simulation at large \cite{Pasik_JCP_1999,Copplestone_IEEE_2017}.
While there is a large body of work on PMLs in general, the issue of
charged particles reaching the PML has attracted relatively little attention.

Historically, solutions to this issue have been proposed and studied
in the context of microwave devices, whereby an electron beam passes
through the simulation domain, and interacts with electromagnetic modes in a cavity
or a waveguide
\cite{Jost_CPC_1997,Pasik_JCP_1999,Ludeking_TOPPJ_2010,Copplestone_IEEE_2017}.
For instance, in \cite{Pasik_JCP_1999}, the authors observed that spurious
electrostatic fields can build up at the boundary of the simulation domain,
if charged particles are simply removed from the simulation as soon as
they enter the PML region.
(Note that, by default, this is how charged particles are treated in many PIC
codes with PMLs.) In order to mitigate this issue, the authors proposed to
apply numerical diffusion to these fields, by using a Marder-type
divergence-cleaning algorithm \cite{Marder_JCP_1987}, which causes
these fields to decay over time. This does prevent the build up of spurious
fields, but is only effective over timescales longer than the characteristic
diffusion time, which is limited by stability constraints \cite{Pasik_JCP_1999}.
Alternatively, instead of removing the particles, some algorithms
\cite{Jost_CPC_1997,Copplestone_IEEE_2017} allow them to propagate into
the PML region, and to use the corresponding
current density as a source term in the PML equations.
In the case of \cite{Copplestone_IEEE_2017}, this is also combined with a
propagative divergence-cleaning algorithm
\cite{Vay_Proc_1996,Vay_PoP_1998,Munz_CPC_2000,Munz_JCP_2000}.
It is observed that this strongly
mitigates the spurious electrostatic fields at the boundary.

In the above-mentioned studies, the authors assess the efficiency of
their respective algorithms by observing the overall behavior of the simulation,
for their particular physics problem, and by verifying that this overall behavior
is physically realistic. However, they do not study the detailed
behavior of the fields associated with individual particles as they exit the simulation.

By contrast, in the present paper, we first focus on the fields of individual
particles, before moving on to more complex, physically-relevant simulations.
We show that, when an individual particle exits the simulation domain, it can
produce a spurious electrostatic field (as observed in
\cite{Pasik_JCP_1999,Copplestone_IEEE_2017}), but also a spurious
electromagnetic radiation (\cref{sec:spurious}). We propose a set of modified PML equations which, for an outgoing particle in normal incidence, suppress both of these effects, and thus allow individual particles to cleanly exit the simulation domain
(\cref{sec:pml} and \cref{sec:single}) -- as if it was propagating in infinite vacuum.
We then apply this new algorithm in physically-relevant simulations
(\cref{sec:physical}).

\section{Spurious effects associated with particles entering the PML}
\label{sec:spurious}

In order to illustrate the numerical issues that arise when particles enter
the PML, we run PIC simulations of a single macroparticle impinging on a PML,
with a relativistic Lorentz factor $\gamma = 10$. The other simulation
parameters, and the overall simulation setup, are summarized in
\cref{sec:simulation-single}.

\begin{figure}
    \includegraphics[width=\linewidth]{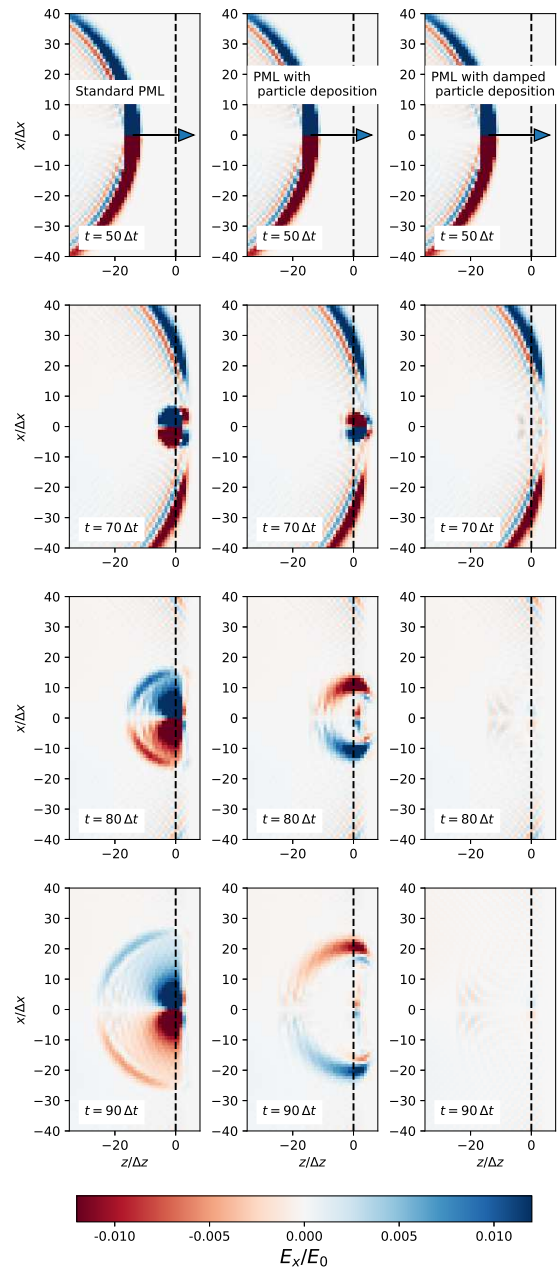}
    \caption{Colormaps of the $E_x$ field, near the edge of the simulation box.
    The PML region corresponds to $z>0$. (In this region, we show the sum of the split
    components $E_x = E_{xy} + E_{xz}$.) The different rows correspond to
    different simulation times, while the different columns correspond to
    different algorithms. Left column: standard PML algorithm
    ($\alpha(z) = 0$ in \cref{eq:MaxwellE}). Middle column: PML with
    current deposition ($\alpha(z) = 1$ in \cref{eq:MaxwellE}).
    Right column: PML with damped particle deposition ($\alpha(z)$ given
    by \cref{eq:alpha}, with $v=c$). The notations $\Delta x$, $\Delta z$ and $\Delta t$
    correspond to the cell sizes and timestep of the simulation respectively.
    The field is normalized by $E_0 \equiv q/(4\pi \epsilon_0 \Delta x^2)$,
    where $q$ is the charge of the macroparticle.
    \label{fig:illustrate}}
\end{figure}

The left column of \cref{fig:illustrate} shows the evolution of the electric
field in the case of the standard PML algorithm. The fields seen in the top
panel are the self-fields associated with the macroparticle. (As
mentioned in \cref{sec:simulation-single}, the macroparticle separated from an
opposite-charge motionless macroparticle at $t=0$. Therefore,
the associated self-fields at $t>0$ have a curved wave-front.) The bottom panels in the left column of \cref{fig:illustrate}
show that two types of spurious fields arise after the macroparticle enters the PML:
\begin{itemize}
    \item a static field, that remains confined close to $z=0$ (domain boundary).
    \item an outward-propagating pulse, that expands in the $z<0$ (interior) region.
\end{itemize}

The existence of these fields can be understood qualitatively. In the standard
PML algorithm, the macroparticle is indeed removed as soon as it enters the PML
region, and thus the current density $\vec{j}$ is zero in the entire simulation box
from then on. In the electromagnetic PIC field update, this is in fact equivalent
to having the macroparticle suddenly stop at the vacuum/PML interface.
(If the macroparticle's velocity $\vec{v}$ is zero, then the associated $\vec{j}$ is
zero as well.) The abovementioned static field can thus be seen as the
space-charge field of this equivalent, motionless macroparticle, while the
outward-propagating pulse can be seen as the electromagnetic radiation
generated by the macroparticle suddenly decelerating to reach a rest state.

Intuitively, one might think that these spurious effects would disappear
if, instead of being removed, the macroparticle is allowed to propagate
and deposit current in the PML. However, the middle column of
\cref{fig:illustrate} shows that this is not the case. Allowing the particle
to deposit current in the PML does remove the spurious static field, but
an outward-propagating spurious pulse still exists. We note that this
remaining pulse might be analogous to the transition radiation associated
with a charged particle crossing the interface between two media with different
electromagnetic properties.

In the next section, we propose a modified PML algorithm that does remove
both the static field and the outward-propagating spurious pulse under
certain conditions, as shown in the right column of \cref{fig:illustrate}.

\section{A modified set of PML equations, for absorption of charged particles}
\label{sec:pml}

\subsection{Continuous equations}
\label{sec:continuous}

In order to describe the proposed algorithm, let us consider that the
interface between the simulation and the PML is at $z=0$, with the PML
lying in the $z>0$ half-space, as represented on \cref{fig:illustrate}.
(Hence, $z$ represents the depth inside the PML for $z>0$.)

We let the particles propagate and deposit their current
density $\vec{j}$ in the PML region ($z>0$) (similarly to
\cite{Jost_CPC_1997,Copplestone_IEEE_2017}), and
use the following split-field Maxwell equations:
\begin{subequations}
    \label{eq:MaxwellE}
    \begin{align}
    \partial_t E_{xy} &= c^2\partial_y B_z \\
    \partial_t E_{xz} &= -c^2\partial_z B_y - \frac{\sigma(z)}{\epsilon_0}E_{xz} - \frac{\alpha(z)}{\epsilon_0}j_x   \\
    \partial_t E_{yz} &= c^2\partial_z B_x - \frac{\sigma(z)}{\epsilon_0}E_{yz} - \frac{\alpha(z)}{\epsilon_0}j_y \\
    \partial_t E_{yx} &= -c^2\partial_x B_z \\
    \partial_t E_z &= c^2(\partial_x B_y - \partial_y B_x) - \frac{\alpha(z)}{\epsilon_0} j_z
    \end{align}
\end{subequations}
\begin{subequations}
    \label{eq:MaxwellB}
    \begin{align}
    \partial_t B_{xy} &= -\partial_y E_z \\
    \partial_t B_{xz} &= \partial_z E_y - \frac{\sigma(z)}{\epsilon_0} B_{xz} \\
    \partial_t B_{yz} &= -\partial_z E_x - \frac{\sigma(z)}{\epsilon_0} B_{yz} \\
    \partial_t B_{yx} &= \partial_x E_z \\
    \partial_t B_z &= -(\partial_x E_y - \partial_y E_x)
    \end{align}
\end{subequations}
\begin{subequations}
    \label{eq:MaxwellSum}
    \begin{align}
    E_{x} &\equiv E_{xy} + E_{xz} \\
    E_{y} &\equiv E_{yz} + E_{yx} \\
    B_{x} &\equiv B_{xy} + B_{xz} \\
    B_{y} &\equiv B_{yz} + B_{yx}
    \end{align}
\end{subequations}
where $\sigma(z)$ is the PML conductivity
(which, in general, varies as a function of the depth $z$ \cite{Taflove}),
and $\alpha(z)$ is a newly-introduced damping coefficient on the current density.
For $\alpha(z)=0$, these equations are identical to the original PML formulation
by B\'erenger \cite{Berenger_JCP_1994}, in which there is no source term
associated with charged particles.
On the other hand, $\alpha(z)=1$ results in a scheme which is similar to
\cite{Jost_CPC_1997,Copplestone_IEEE_2017}.
(As mentioned previously, \cite{Copplestone_IEEE_2017} also adds a
propagative divergence-cleaning correction to these equations.)

By contrast, here we choose the following prescription:
\begin{equation}
\label{eq:alpha}
\alpha(z) = \exp\left( -\int_0^z \frac{\sigma(z')}{\epsilon_0} \frac{dz'}{v} \right)
\end{equation}
where $v$ is the assumed velocity of the exiting particles.
(This assumed velocity is discussed in more detail in \cref{sec:limitations}.)
Notice that the value of the damping coefficient $\alpha(z)$ decreases
monotonically from 1 at the PML interface ($z=0$) towards 0 deep inside the PML.

In \cref{sec:demonstration}, we show that, for a particle
\emph{in normal incidence with a velocity $v$},
these modified PML equations (with $\alpha(z)$ given by \cref{eq:alpha})
constitute an \emph{ideal} open boundary.
More specifically, when using these equations \emph{in the PML domain},
the fields associated with the particle \emph{in the physical domain} are
\emph{exactly} the same as if the PML were replaced by an infinite vacuum region.
In particular, there is no spurious electrostatic field and electromagnetic
radiation as the particle transitions from the physical domain to the PML region.

We note that this proposed PML algorithm has an intuitive interpretation.
One indeed expects the fields associated with the particle to be
progressively damped in the PML. However, this cannot consistently occur if the
source of these fields (the current $\vec{j}$) is not damped as well.
In this sense, it seems reasonable that the damping of $\vec{j}$ needs to
be ``matched'' to the natural damping rate of the field in the PML.
However, the exact matching condition itself (given by \cref{eq:alpha})
is less intuitive, and requires a proper derivation -- as given in \cref{sec:demonstration}.

Finally, note that the extension of \cref{eq:MaxwellE,eq:MaxwellB,eq:MaxwellSum} to 2D
(or 1D) Cartesian geometries can be readily obtained by setting derivatives
term to 0 for the corresponding invariant directions of space.

\subsection{Discretization}

The above equations and considerations are valid in the continuous limit.
However, in a PIC algorithm, \cref{eq:MaxwellE,eq:MaxwellB,eq:MaxwellSum} would
of course be discretized in time and space. For that purpose, notice that the
newly-introduced terms of the form $\alpha(z)j_x$, $\alpha(z)j_y$, $\alpha(z)j_z$
do not affect the usual finite-difference time-domain (FDTD) discretization of
the other terms in \cref{eq:MaxwellE,eq:MaxwellB,eq:MaxwellSum}, on a staggered
Yee grid \cite{Yee1966}. In addition, these newly-introduced terms are naturally
properly centered on a Yee grid, so that no additional
interpolation or averaging is needed. Finally, in the discretized system,
$\alpha(z)$ will be evaluated at the nodes and cell-centers of the staggered
grid. If $\sigma(z)$ is chosen to be a simple, analytical function (as is
often the case; see \textit{e.g.} \cref{sec:simulation-single}), then $\alpha(z)$ can
readily be expressed in a closed analytical form from \cref{eq:alpha} before being
evaluated at these points.

As a result of the discretization, the ideal absorbing properties
of the continuous system \cref{eq:MaxwellE,eq:MaxwellB,eq:MaxwellSum,eq:alpha}
may be slightly affected. (Note that the original PMLs themselves were
also derived in the continuous limit \cite{Berenger_JCP_1994}, and that their
absorption properties become imperfect once discretized.)
While the bottom right panel of \cref{fig:illustrate} does show some faint
remaining spurious fields that may indeed be due to the discretization,
it nonetheless represents a major improvement compared to the other methods
(middle and left panels).

\subsection{Discussion and limitations}
\label{sec:limitations}

As mentioned in \cref{sec:continuous,sec:demonstration},
the proposed PML scheme is well-adapted for macroparticles entering the PML in
normal incidence, and with a velocity that matches the assumed velocity $v$
in \cref{eq:alpha}. On the other hand, if a macroparticle is incident with an
oblique angle, or with a velocity different than the assumed one, the
absorption efficiency of the proposed PML will be lower.
This will be examined and quantified in \cref{sec:single}.

We note that, while the limitation on the angle cannot be easily overcome,
the limitation on the velocity can be lifted by a slightly more complex scheme.
Rather than damping the current density uniformly by $\alpha(z)$ for all
particles, we can set $\alpha(z)=1$ in \cref{eq:MaxwellE}
and instead damp the \emph{weight} of each macroparticle according to:
\[ w_i' = \tilde{\alpha}_i(t) w_i \qquad
\tilde{\alpha}_i(t) = \exp\left(-\int_{t_{i}}^{t}
\frac{\sigma(z_i(t'))}{\epsilon_0}dt'\right) \]
where $w_i$ is the original weight of macroparticle $i$, $t_{i}$ is the
time at which the macroparticle enters the PML, and $z_i(t)$ is the
position of that macroparticle as a function of time.
Due to the linearity of the Maxwell equations, the absorption property of
this alternative scheme \emph{for the self-fields of each individual
macroparticle} is identical to that of the original scheme, where $v$
in \cref{eq:alpha} would match the velocity of the individual macroparticles.
For the sake of simplicity, this alternative scheme is not discussed further
in the rest of this manuscript.

Importantly, while the proposed scheme constitutes an ideal open boundary
for a particle being \emph{absorbed} in the PML, this property does not
necessarily carry over for a particle being \emph{emitted} from the PML.
More precisely, when using \cref{eq:alpha} with $v>0$, the PML will efficiently
absorb the fields associated with
a particle propagating with $+v$ into the PML, but not necessarily the fieds
associated with a particle initialized inside the PML and propagating with
$-v$ into the simulation domain. Fundamentally this is because, unlike
the Maxwell equations, the PML equations are not time-reversible.


\section{Single-particle tests}
\label{sec:single}

In this section, we quantify the impact of the incidence angle and velocity
of the incident particle on the absorption properties of the PML.
To this end, we ran additional simulations of a macroparticle crossing the PML,
similarly to the case represented in \cref{fig:illustrate},
while varying the angle and velocity of that macroparticle.
(The other simulation parameters are the same as, in
\cref{sec:simulation-single}.) In each case, we compared the results with a
reference simulation in which the simulation box is extended in the $z>0$ region,
so that in this case the particle does not exit the box during the course of the simulation.
(In practice, we double the size of the simulation box along the $z$ direction,
compared to the base case described in \cref{sec:simulation-single}.)
In this reference simulation, the particle self-fields are therefore free
of the numerical artifacts associated with a particle crossing the PML interface.

We then define the relative error as:
\begin{equation}
\mathcal{E} = \frac{\int \!d\vec{x} \;
\left[(\vec{E}-\vec{E}_{ref})^2 + c^2(\vec{B}-\vec{B}_{ref})^2\right]}
{\int \!d\vec{x} \;\left[\vec{E}_{ref}^2 + c^2\vec{B}_{ref}^2\right] }
\label{eq:error}
\end{equation}
where we integrate over the physical cells of the original simulation.
We evaluate this quantity
at the end of the simulation i.e., $t = 125 \Delta t$, while the macroparticle
crosses the PML interface at $t = 65 \Delta t$.
Note that, because of the presence of a motionless opposite-charge particle
in the simulation box (see \cref{sec:simulation-single}) the denominator in \cref{eq:error}
is non-zero.

\subsection{Impact of the incidence angle}
\label{sec:angle}

\begin{figure}
    \includegraphics[width=\linewidth]{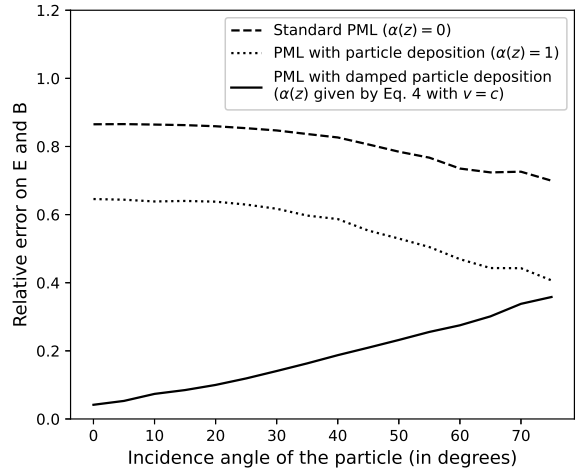}
    \caption{Relative error on $E$ and $B$ due to numerical artifacts
    at the PML interface (as defined in \cref{eq:error}) as a function of
    the incidence angle of the macroparticle, for different PML schemes.
    \label{fig:angle_error}}
\end{figure}

We first vary the incidence angle of the particle from $0^{\circ}$
(normal incidence) to $75^{\circ}$, while keeping
the velocity of the particle fixed (corresponding to a Lorentz factor
$\gamma = 10$). The corresponding relative error for different PML schemes
is shown in \cref{fig:angle_error}. The different PML schemes all implement
\cref{eq:MaxwellE,eq:MaxwellB,eq:MaxwellSum}, but use different choices for
$\alpha(z)$, as indicated in the legend of \cref{fig:angle_error}.

As can be seen in \cref{fig:angle_error}, the error is high for the standard
PML scheme ($\alpha(z)=0$) and, to a lower extent, for the PML scheme with
undamped particle deposition ($\alpha(z)=1$). This is because of the presence
of spurious fields at the PML interface, as described in \cref{sec:spurious}.
In the case of the damped particle deposition ($\alpha(z)$ given by
\cref{eq:alpha} with $v=c$), the error is close to zero for normal incidence,
thereby confirming the predictions of \cref{sec:pml}. As expected, the error
grows when the incidence angle increases. However, it is worth noting
that, even at high incidence angle, this error remains lower than that of the other
PML schemes.

\subsection{Impact of the incident velocity}

\begin{figure}
    \includegraphics[width=\linewidth]{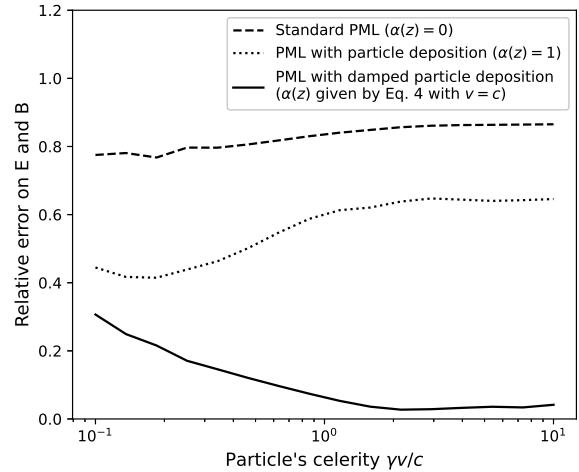}
    \caption{Relative error on $E$ and $B$ due to numerical artifacts
    at the PML interface (as defined in \cref{eq:error}) as a function of
    the celerity of the macroparticle, for different PML schemes.
    \label{fig:velocity}}
\end{figure}

We then varied the incident velocity of the macroparticle, from non-relativistic
($v \ll c$) to ultra-relativistic, in normal incidence. (For low $v$, the macroparticle is initialized
closer to the PML, so that the particle still crosses the PML interface
at the same time $t=65 \Delta t$.) We compare the different PML schemes in
\cref{fig:velocity}. In the case of the damped PML, we use \cref{eq:alpha}
with a fixed $v$ ($v=c$ in this case).

Again, \cref{fig:velocity} shows that the error is relatively high for
the standard PML scheme ($\alpha(z)=0$) and the PML with undamped particle
deposition ($\alpha(z)=1$). As expected, for low velocity ($v<c$), the PML
scheme with damped particle deposition also has a significant error,
due to the fact that we use $v=c$ in \cref{eq:alpha}.
(Recall however from \cref{sec:limitations} that this limitation could be overcome,
even for multiple macroparticles with different velocities, by a modified scheme where the macroparticle weight is progressively damped.)
Yet for relativistic
velocities ($\gamma v/c > 1$), the error in \cref{fig:velocity} becomes very low.
This also indicates that, in simulated scenario where most exiting particles
are known to be relativistic, setting $v=c$ independently of the actual
energy of the particles is a reasonable choice.

\section{Application in accelerator simulations}
\label{sec:physical}

In this section, the new PML scheme is tested with two physically-relevant
scenarios.

\subsection{Beam passing through a metallic cavity}
\label{sec:cavity}

\begin{figure*}
    \includegraphics[width=\textwidth]{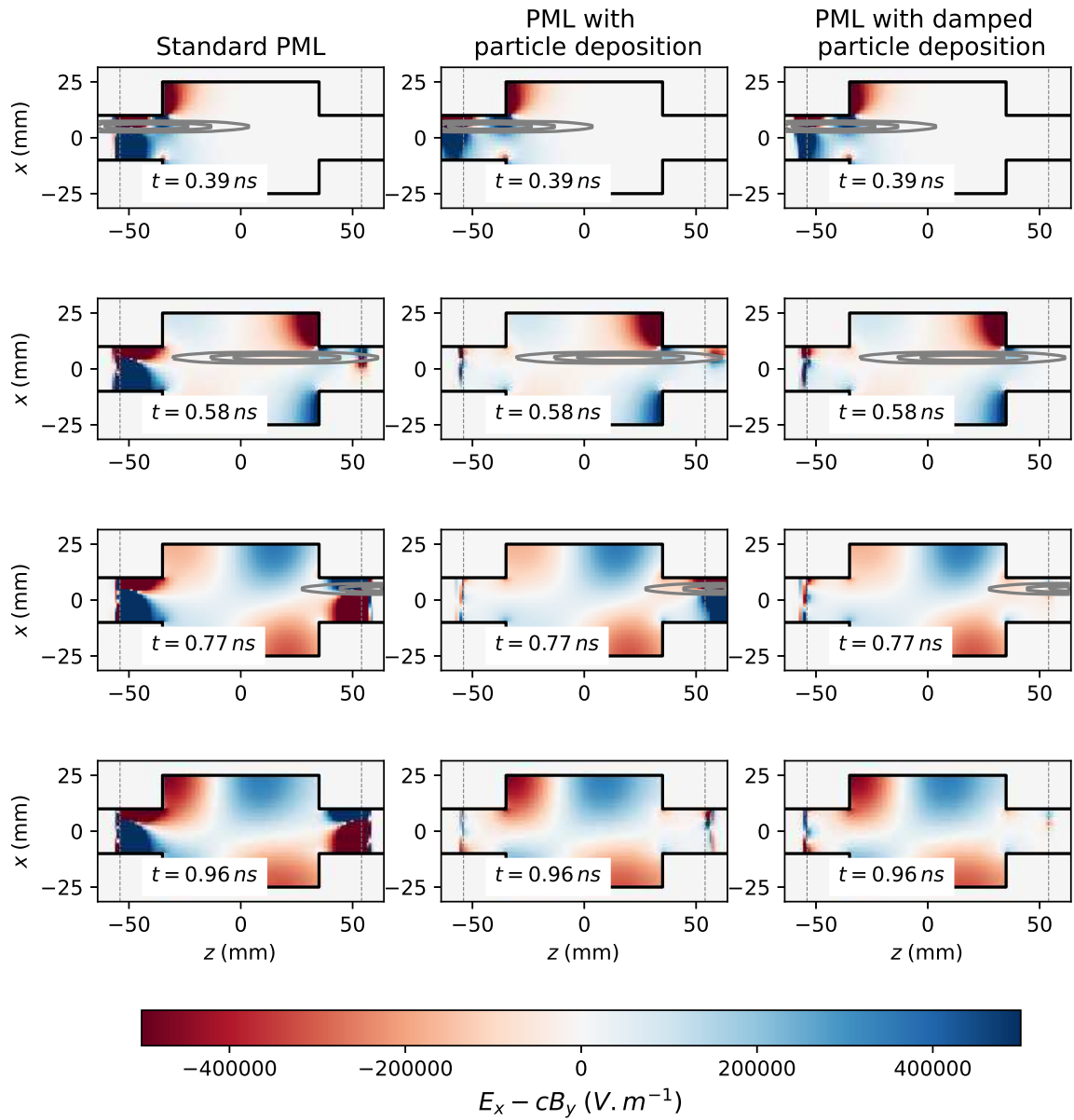}
    \caption{Colormaps of $E_x - cB_y$ in the simulation box.
    The dashed lines represent the end of the actual simulation box, and the beginning of the PML cells.
    (In the PML cells, we show the sum of the split components for $E_x$ and $B_y$.)
    The black lines represent the cavity boundary, while the gray
    contour lines represent the proton beam density.
    The different rows correspond to different simulation times,
    while the different columns correspond to different algorithms.
    Left column: standard PML algorithm
    ($\alpha(z) = 0$ in \cref{eq:MaxwellE}). Middle column: PML with
    current deposition ($\alpha(z) = 1$ in \cref{eq:MaxwellE}).
    Right column: PML with damped particle deposition ($\alpha(z)$ given
    by \cref{eq:alpha}, with $v=c$).
    \label{fig:cavity}}
\end{figure*}

We consider a proton beam passing through a simplified,
rectangular accelerator cavity.
Since the proton beam
is relatively long, it is continuously injected on one side of the simulation
box and then absorbed on the other side. Both sides of the simulation box
are terminated by PMLs. The simulation parameters are summarized
in \cref{sec:simulation-cavity}. Note that, in principle, the same
configuration could also be used to simulate a train of proton bunches
going through the cavity, instead of a single bunch.

Figure \ref{fig:cavity} shows colormaps of the quantity $E_x - cB_y$ at
different times and with different PML schemes. Note that $E_x - cB_y$
corresponds, to a good approximation, to the transverse force felt by
ultra-relativistic particles traveling in the $+z$ direction.
Note also that the space charge field of the (relativistic) proton beam
largely cancels out when evaluating $E_x - cB_y$. 

Of particular interest is the behavior of the fields at
both ends of the simulation box ($z=\pm 50\,\mathrm{mm}$).
As expected, the standard PML scheme leads to a strong spurious field
(left column in \cref{fig:cavity}).
At the left-hand side of the simulation box ($z=-50\,\mathrm{mm}$),
this issue is significantly reduced by depositing the particle current in
the PML (middle and right column), but is not completely removed.
This is particularly visible at $t = 0.39\,\mathrm{ns}$. In the case
of the proposed PML scheme (right column in \cref{fig:cavity}), this remaining
spurious field is indeed expected from the discussion of
\cref{sec:limitations}: although the proposed PML scheme is well-adapted for
exiting particles, it is not as efficient for particles entering the simulation
 box from the PMLs.

By contrast, the advantage of the proposed PML scheme is clearly visible
at the right-hand side of the simulation box ($z=+50\,\mathrm{mm}$),
in particular at $t = 0.77\,\mathrm{ns}$. In this case, simply depositing
the particle current in the PML without damping still leads to significant
spurious fields at the right-side boundary  (middle column). By contrast, the proposed
PML scheme largely removes these spurious fields (right column).

\begin{figure}
\includegraphics[width=\linewidth]{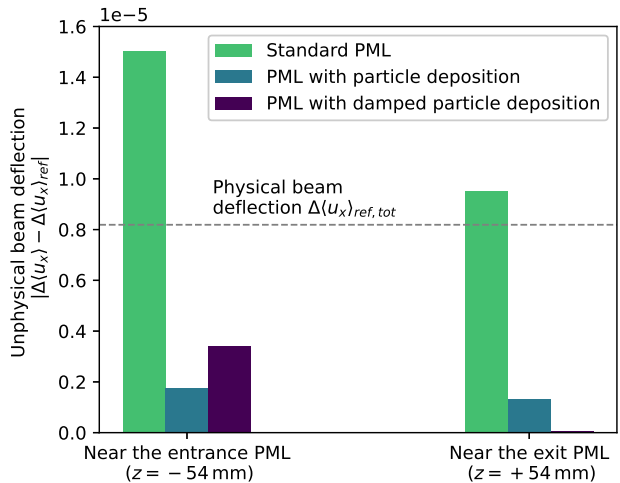}
\caption{Beam deflection (quantified by the change in the mean dimensionless momentum of the beam $\langle u_x \rangle = \langle p_x \rangle /mc$) for the cases represented in Fig.~\ref{fig:cavity}. The unphysical deflection imprinted on the beam near the left and right PMLs (colored bars) is compared with the total physical deflection from the entire cavity (dashed line). See the main text for details on the calculation of these different quantities.\label{fig:beamdeflection}}
\end{figure}

In order to confirm the advantage of the proposed scheme, we now quantify the impact of the spurious fields on the beam itself. Note that, since the centroid position of the beam in $x$ is offset with respect to the axis of the cavity (see Fig.~\ref{fig:cavity}, and appendix \ref{sec:simulation-cavity}), the $E_x - cB_y$ field in the cavity results in a net deflection of the beam along $x$. We quantify this deflection by computing the mean of the dimensionless momentum along $x$ i.e. $\langle u_x \rangle \equiv \langle p_x \rangle / mc$. The physical deflection due to the cavity is obtained by running a reference simulation, whereby the box is made twice larger in the $z$ direction, and by recording the change in $\langle u_x \rangle$ between $z = -54\,\mathrm{mm}$ and $z = +54 \,\mathrm{mm}$. The corresponding value is shown as a dashed black line in Fig.~\ref{fig:beamdeflection}.

However, in addition to this physical deflection, the spurious fields seen near the PMLs in Fig.~\ref{fig:cavity} also imprint an \emph{unphysical} deflection on the beam. This unphysical deflection is evaluated by calculating the deflection $\Delta \langle u_x \rangle$ between $z=-54\,\mathrm{mm}$ and $z=-35\,\mathrm{mm}$ (for the left PML) and $z=+35\,\mathrm{mm}$ and $z=+54\,\mathrm{mm}$ (for the right PML), and by subtracting the corresponding deflection $\Delta \langle u_x \rangle_{ref}$ from the reference simulation in those same sections of the simulation. The corresponding values, for the different PML algorithms tested in Fig.~\ref{fig:cavity}, are also shown in Fig.~\ref{fig:beamdeflection}.

As can be seen, for the standard PML, the unphysical deflection at the left and right PMLs (green bars) both exceeds the total physical deflection from the whole cavity (black dashed line). As expected, the PML with undamped current deposition significantly improves this, but the unphysical deflection (blue bars) is still a non-negligible fraction of the total physical deflection. Finally, the PML with damped deposition (purple bar) is seen to result in a negligible unphysical deflection \emph{at the right PML}, while it still results in a significant unphysical deflection at the left PML (purple bars). This is consistent with the observations of Fig.~\ref{fig:cavity}, and with the fact that the proposed PML algorithm is well-adapted for exiting particles, but not for entering particles. We note that, in principle, one could use the undamped PML algorithm on the left PML, and the damped PML algorithm on the right PML in order to minimize the total spurious effect on the beam.

\subsection{Laser-wakefield acceleration}
\label{sec:lwfa}

\begin{figure*}
    \includegraphics[width=0.95\textwidth]{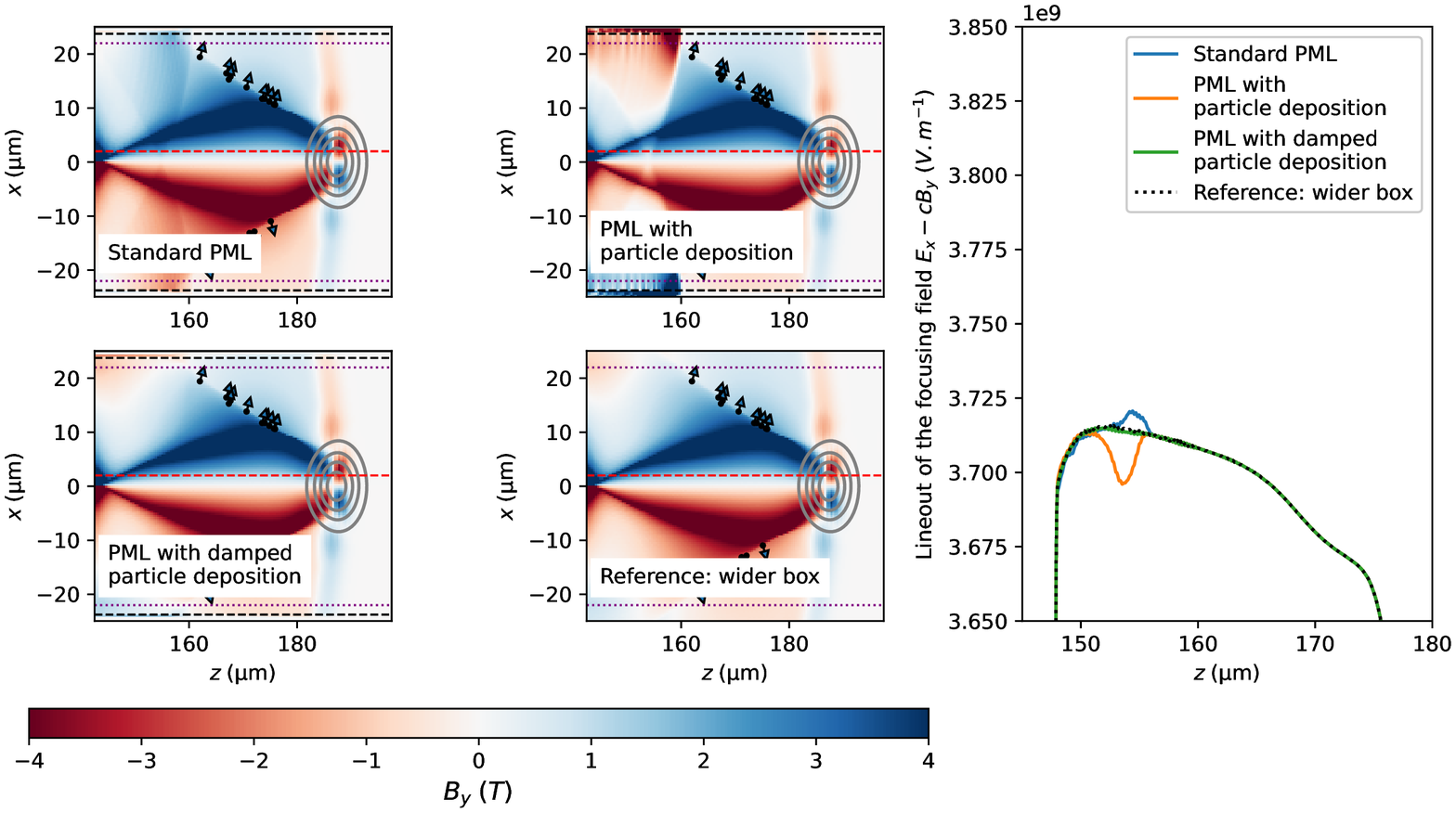}
    \includegraphics[width=0.95\textwidth]{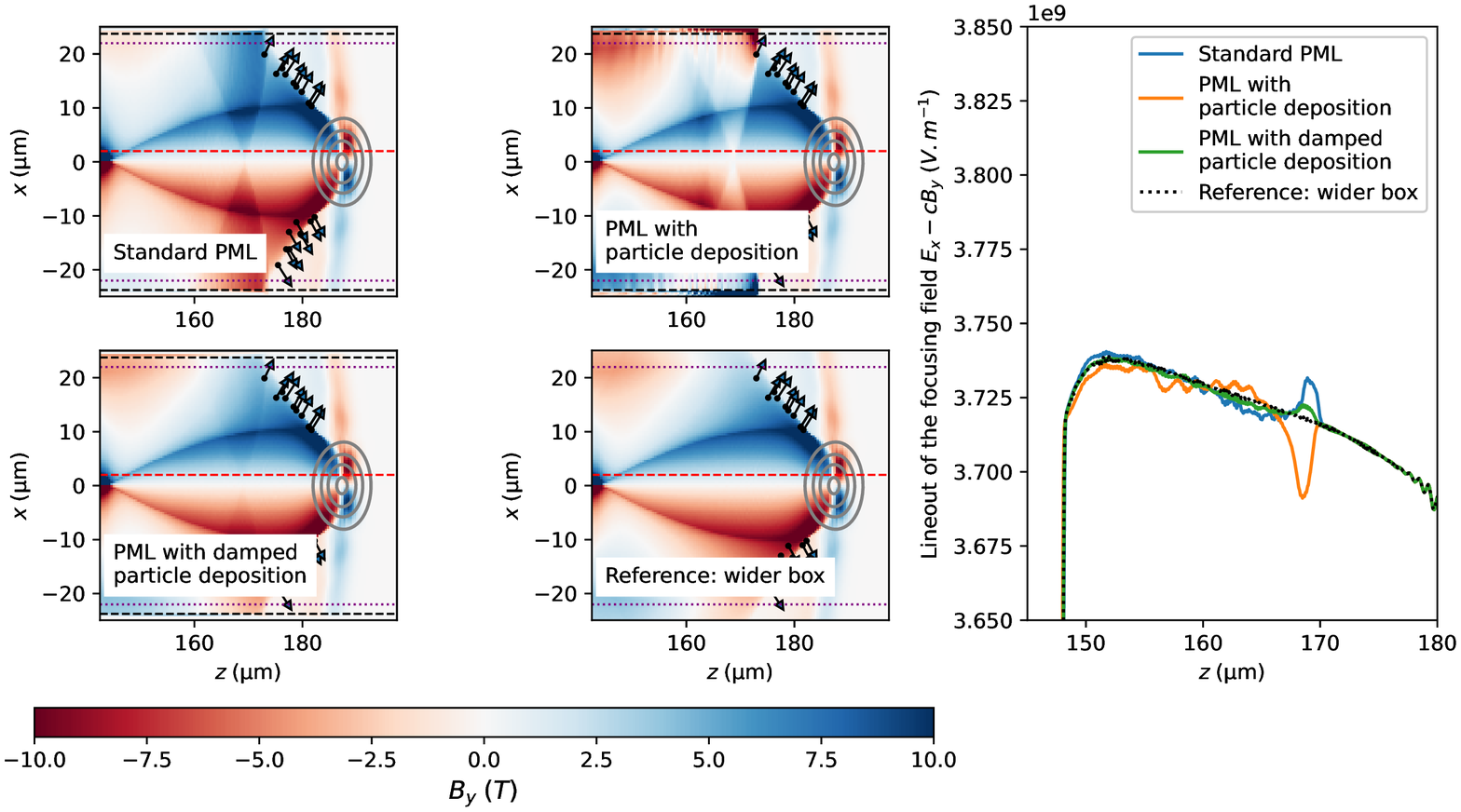}
    \caption{\emph{Left:} Colormaps of the transverse magnetic field $B_y$ in the simulation box.
    The arrows represent the velocity vectors of a sample of the plasma electrons that were expelled by the laser driver.
    The gray contour lines indicate the position of the laser driver itself.
    The black dashed lines correspond to the transverse boundaries of the actual simulation box, and the beginning of the PML cells.
    (In the PML cells, we show the sum of the split components of $B_y$.)
    The purple dotted lines represent the transverse limits of the plasma that is injected in front of the laser beam in the simulation.
    (As is often the case in this type of simulation, some amount of vacuum was left between the PML and the plasma.)
    The red dashed line corresponds to the position of the lineout shown on the right. 
    The different colormaps correspond to different PML algorithms, as well as to a reference case (rightmost column) where the simulation box was two times wider in $x$ and $y$. (The colormap does not show the full simulation box in this case.)
    \emph{Right:} Lineouts of the focusing field $E_x - cB_y$, close to the axis, for the different PML algorithms.
    The upper and lower halves of this figure correspond to laser amplitude values of $a_0=2$ and $a_0=3$, respectively. 
    \label{fig:LWFA}}
\end{figure*}


We now consider simulations of a laser-driven plasma accelerator \cite{EsareyRMP2009} in the blow-out regime \cite{Pukhov_2004}.
In the simulations that we performed, an intense ultra-short laser pulse propagates through an underdense plasma, and drives a non-linear plasma wakefield. A moving window is used
to follow the pulse as it propagates. Snapshots of the simulations are shown in \cref{fig:LWFA} for two different values of the laser amplitude: $a_0=2$ (top) and $a_0=3$ (bottom). The other simulation parameters are summarized in \cref{sec:simulation-lwfa}.

In this configuration, it is common for the head of the laser pulse to progressively diffract, and thus to reach the transverse boundaries of the simulation box, in grazing incidence. Therefore, PMLs are often used as transverse boundary conditions, so as to prevent the diffracting laser field from spuriously reflecting back into the simulation box.

At the same time, in the blow-out regime, a fraction of the plasma electrons also reach the transverse boundary, after being violently expelled by the ponderomotive force of the laser driver. A sample of these plasma electrons are represented for instance in \cref{fig:LWFA}, along with their velocity vector. As can be seen, these plasma electrons are concentrated in a well-defined stream, which impacts the transverse boundary at $z \simeq 160$ (upper half) or $z \simeq 170\,\mathrm{\mu m}$ (lower half). As these electrons enter the PML region, they generate a spurious electromagnetic radiation which reflects back towards the axis, as seen again in \cref{fig:LWFA}. For the standard PML scheme (upper left colormap) and the PML scheme with particle deposition without damping (upper right colormap), this spurious field is quite strong, and it distorts the structure of the fields inside the plasma bubble. This distortion is particularly visible in the lineouts of $E_x - cB_y$, on the right side of Fig.~\cref{fig:LWFA}. This quantity corresponds to the focusing field that a relativistic electron beam would experience, if it was co-propagating inside the plasma bubble. Therefore, any distortion of this focusing field can have implications for the preservation of ultra-low beam emittance, and for the potential development of a hosing instability, in a laser-wakefield accelerator.

By contrast, when damping the deposited current in the PML (lower left colormap), this spurious distortion is either practically suppressed (upper half in \cref{fig:LWFA}, corresponding to $a_0 = 2$) or at least significantly mitigated (lower half in \cref{fig:LWFA}, corresponding to $a_0=3$). 
This can be seen in particular by comparing this case with the reference case (lower right colormap), in which the simulation box is two times wider in $x$ and $y$ (compared with the parameters of \cref{sec:simulation-lwfa}) and thus avoids the artifacts associated with jets of electrons crossing the PML interface.  
The fact that the proposed PML scheme is less efficient for $a_0 = 3$ (bottom half in Fig.~\cref{fig:LWFA}) is most likely due to the fact that the electrons are impinging on the PML at a more oblique angle in this case, as can be seen from the velocity vectors in \cref{fig:LWFA}. (Recall from \cref{sec:angle} that the proposed PML scheme is most efficient for normally incident particles.) This motivates further development of this PML scheme in the future, so as to make it more robust to the incidence angle of the particles.

\section{Conclusion}

In summary, in this paper we proposed a new PML scheme, whereby macroparticles
deposit their current into the PML cells and whereby this current is damped
by an optimal coefficient (\cref{eq:alpha}) which minimizes numerical artifacts.
It was shown, mathematically and through single-particle simulations, that
this PML scheme behaves practically as an ideal \emph{vacuum} open boundary condition, for macroparticles entering the PML in normal incidence and with a known velocity.
In particular, the proposed scheme suppresses both the spurious static field
and spurious radiation that is usually associated with particles exiting the
simulation box.

This scheme could have applications in various physical
scenarios, including plasma simulations with 
out-flowing relativistic jets of particles as well as simulations of trains of particle bunches traveling through complex structures. Moreover, in the future,
the scheme could be applied in the context of mesh refinement,
for algorithms that involve refined patches surrounded by PML\cite{Vay2004}.

\section*{DATA AVAILABILITY}
The data that supports the findings of this study is  openly available in Zenodo at
\url{https://doi.org/10.5281/zenodo.7011364}.

\begin{acknowledgments}
The authors acknowledge the discussions and code contributions from
the whole WarpX development team, that made this work possible.
This work was partly supported by the Director, Office of Science,
Office of High Energy Physics, U.S. Dept. of Energy under Contract
No. DE-AC02-05CH11231, as well as by the Exascale Computing Project (17-SC-20-SC),
a collaborative effort of two U.S. Department of Energy organizations
(Office of Science and the National Nuclear Security Administration).

This research used resources of the National Energy
Research Scientific Computing Center, a DOE Office of Science User
Facility supported by the Office of Science of the U.S. Department of
Energy under Contract No. DE-AC02-05CH11231, and of the Oak Ridge Leadership
Computing Facility at the Oak Ridge National Laboratory, which is supported
by the Office of Science of the U.S. Department of Energy under Contract No. DE-AC05-00OR22725.
\end{acknowledgments}

\appendix
\section{Simulation parameters}

This section lists the parameters that were used in the different simulations mentioned in the main text. All simulations were run with the open-source PIC code WarpX \cite{warpx}, available at \url{https://github.com/ECP-WarpX/WarpX}.

\subsection{Single-particle simulations}
\label{sec:simulation-single}

In the single-particle tests, a charged particle is initialized close to the middle of the simulation box,
with a velocity directed towards the PML. The $\boldsymbol{E}$
and $\boldsymbol{B}$ fields are initialized to zero. In order to ensure
$\boldsymbol{\nabla}\cdot\boldsymbol{E} = \rho/\epsilon_0$,
a motionless particle of opposite charge is placed at the initial position
of the moving particle. This motionless opposite-charge particle therefore produces an
associated space charge field in the simulation box, but its amplitude is low at
the position of the PML.

The simulations use the finite-difference Cole-Karkkainen (CK) Maxwell solver
\cite{Vay_JCP_2011} with cubic cells ($\Delta x = \Delta y = \Delta z$), at the
Courant-Friedrichs-Lewy limit ($c\Delta t = \Delta z$).
The current deposition operation uses the charge-conserving Esirkepov scheme
\cite{Esirkepov_CPC_2001}, with cubic particle shape factors
(also known as Piecewise Cubic Spline interpolation). An additional binomial filter
\cite{Birdsall2004} is applied to the deposited current before updating the fields.

The simulation domain consists of a 3D box of $128\times128\times128$ cells,
surrounded by PMLs extending 8 cells beyond the boundary.
The PML conductivity $\sigma(z)$ is given by
\begin{equation}
\label{eq:conductivity}
\sigma(z) = \frac{4\epsilon_0 c}{\Delta z} \left(\frac{z}{6\Delta z}\right)^2
\end{equation}

\subsection{Simulations of a rectangular cavity}
\label{sec:simulation-cavity}

The simulation of \cref{sec:cavity} uses the Yee solver \cite{Yee1966} with cubic cells
($\Delta x = \Delta y = \Delta z = 1\,\mathrm{mm}$), at the
Courant-Friedrichs-Lewy limit ($c\Delta t = \Delta z/\sqrt{3}$).
The cavity is modeled by setting the $\vec{E}$ and $\vec{B}$ fields
to zero in the cells covered by the cavity.
The current deposition operation uses the charge-conserving Esirkepov scheme
\cite{Esirkepov_CPC_2001}, with linear particle shape factors and
an additional binomial filter \cite{Birdsall2004}. The simulation box
consists of 64$\times$64$\times$108 cells, with PMLs extending over
another 10 cells on both sides, in the $z$ direction. The conductivity
in the PMLs is given by:
\begin{equation}
\sigma(z) = \frac{4\epsilon_0 c}{\Delta z} \left(\frac{z}{10\Delta z}\right)^2
\end{equation}
The proton beam consists of 40 nC and is monoenergetic with a Lorentz factor
$\gamma = 479$. It has a Gaussian spatial distribution with a 18.8 mm longitudinal RMS size, a 0.2 mm transverse RMS size, and a +5 mm offset in the $x$ direction.

\subsection{Laser-wakefield acceleration simulations}
\label{sec:simulation-lwfa}

The simulation of \cref{sec:lwfa} uses the Cole-Karkkainen (CK) Maxwell solver
\cite{Vay_JCP_2011} with
($\Delta x = \Delta y = 0.125 \,\mathrm{\mu m}$ and $\Delta z = 0.025\,\mathrm{\mu m}$), at the
Courant-Friedrichs-Lewy limit ($c\Delta t = \Delta z$).
The current deposition operation uses the charge-conserving Esirkepov scheme
\cite{Esirkepov_CPC_2001}, with cubic particle shape factors (also known as Piecewise Cubic Spline interpolation) 
and an additional binomial filter \cite{Birdsall2004}. The simulation box
consists of 380$\times$380$\times$2200 cells, with PMLs extending over
another 10 cells at the $x$ and $y$ boundaries. The conductivity
in the PMLs is given by \cref{eq:conductivity}.

The laser driver has a Gaussian intensity profile with a 6 $\mathrm{\mu m}$ waist and a 15 fs duration, and is polarized along the $y$ axis. The background plasma has a density of $2\times 10^{17}\, \mathrm{cm^{-3}}$ and is represented with 1 macroparticle per cell.

\section{Derivation of the damping rate $\alpha(z)$ that removes spurious
effects for a particle in normal incidence}
\label{sec:demonstration}

\subsection{Statement of the problem}

In this section, we consider a charged particle propagating along $z$
(normal incidence) at constant speed $v$, and crossing the vacuum/PML interface
($z=0$) at $t=0$. In particular, the current density $\vec{j}$ that contributes
to the regular Maxwell equations ($z<0$) and PML equations ($z>0$) is:
\begin{equation}
\label{eq:singleparticlecurrent}
\vecthree{j_x}{j_y}{j_z} = q\;\delta(x, y, z-vt)\vecthree{0}{0}{v}
\end{equation}

In these conditions, \emph{if there are no spurious effects at the vacuum/PML
interface}, we expect the fields in
the physical domain $(z<0)$ to be those of a relativistic particle in
infinite vacuum \cite{Jackson}:
\begin{subequations}
\label{eq:physicalfield}
\begin{align}
\vecthree{E_x}{E_y}{E_z} = \frac{\gamma q}{4\pi \epsilon_0} \frac{1}{{r'}^{3}}\vecthree{x}{y}{z-vt}  \qquad \forall z<0\\
\vecthree{B_x}{B_y}{B_z} = \frac{\gamma v q}{4\pi \epsilon_0 c^2} \frac{1}{{r'}^{3}}\vecthree{-y}{x}{0} \qquad \forall z<0
\end{align}
\end{subequations}
with $r' = [x^2 + y^2 + \gamma^2(z-vt)^2]^{1/2}$.
In general, however, the fields in the physical domain do \emph{not} correspond
to \cref{eq:physicalfield}, because of spurious reflections
at the vacuum/PML interface, as seen in \cref{fig:illustrate}.

A necessary and sufficient condition for the absence of reflections at the
vacuum/PML interface is that the expression of $\vec{E}$ and $\vec{B}$
\emph{inside the PML} is continuous with the incident field
\cref{eq:physicalfield} at $z=0$. (Note that a similar reasoning on continuity
is used when calculating the reflection of plane waves
at the interface between two dielectrics, in classical electromagnetics \cite{Jackson}.
In particular, the case where the transmitted wave is continuous with the
incident wave corresponds to Brewster's angle, and in this case the amplitude of
the reflected wave is zero.)

This condition of continuity would be satisfied, for instance, for the following
expression of $\vec{E}$ and $\vec{B}$ \emph{inside the PML}
\begin{subequations}
\label{eq:PMLfield}
\begin{align}
\vecthree{E_x}{E_y}{E_z} = \alpha(z)\frac{\gamma q}{4\pi \epsilon_0}
 \frac{1}{{r'}^{3}}\vecthree{x}{y}{z-vt}  \qquad \forall z>0\\
\vecthree{B_x}{B_y}{B_z} = \alpha(z)\frac{\gamma v q}{4\pi \epsilon_0 c^2}
 \frac{1}{{r'}^{3}}\vecthree{-y}{x}{0} \qquad \forall z>0
\end{align}
\end{subequations}
under the condition
\begin{equation}
\label{eq:continuitycondition}
\alpha(z=0) = 1.
\end{equation}
The expression in \cref{eq:PMLfield} is motivated by physical intuition:
if the current density is damped by a factor $\alpha(z)$ in the PML equations
\cref{eq:MaxwellE}, then we expect the associated fields to
be damped proportionally. There is however no guarantee at this point that
the expression of the fields \cref{eq:PMLfield} is indeed a mathematical
solution of the PML equations
\cref{eq:MaxwellE,eq:MaxwellB,eq:MaxwellSum}.
(By contrast, the expression of the fields in the physical domain
\cref{eq:physicalfield} are known to be a solution of the
regular Maxwell equations.)

In the rest of this appendix (\cref{sec:FourierFormulation,sec:alphaCondition}),
we search for the conditions on $\alpha(z)$
under which \cref{eq:PMLfield} is indeed a solution of
the PML equations \cref{eq:MaxwellE,eq:MaxwellB,eq:MaxwellSum}, and show
that this happens only when $\alpha(z)$ has
the expression given by \cref{eq:alpha}.

Notice incidentally that this appendix does \emph{not} show that
\cref{eq:alpha} is the \emph{only} solution that suppresses spurious effects
at the PML interface, since there could be expressions of $\alpha(z)$ that
suppress spurious effects when applied in \cref{eq:MaxwellE,eq:MaxwellB,eq:MaxwellSum},
without $\vec{E}$ and $\vec{B}$ necessarily
having the form given by \cref{eq:PMLfield}, in the PML.

\subsection{Formulation in Fourier space}
\label{sec:FourierFormulation}

Here we search for the conditions under which the Ansatz fields \cref{eq:PMLfield}
satisfy the PML equations \cref{eq:MaxwellE,eq:MaxwellB,eq:MaxwellSum}.
We note that this is equivalent to searching the conditions under which the Fourier
transform of \cref{eq:PMLfield} satisfies the Fourier transform of
\cref{eq:MaxwellE,eq:MaxwellB,eq:MaxwellSum}. The formulation
in Fourier space is preferred here, since it simplifies calculations and avoids
the need to deal with derivatives and singularities at the exact position
of the particle.

Here we use the following definition for the Fourier transform in $x$, $y$ and $t$:
\begin{align}
\sp{F}(k_x, &k_y, z, \omega) = \nonumber\\
&\int dt\, dx\, dy\;e^{-ik_x x -ik_y y + i\omega t} F(x, y, z, t)\\
F(x, &y, z, t) = \nonumber\\
&\frac{1}{(2\pi)^{3}}\int d\omega\, dk_x\, dk_y\;e^{ik_x x + ik_y y- i\omega t} \sp{F}(k_x, k_y, z, \omega)
\end{align}
With this definition, the Fourier transform of the PML equations
\cref{eq:MaxwellE,eq:MaxwellB,eq:MaxwellSum} (with $\vec{j}$ given by the
single-particle expression \cref{eq:singleparticlecurrent}) is:
\begin{subequations}
    \label{eq:FourierMaxwellE}
    \begin{align}
    -i\omega \sp{E}_{xy} &= c^2ik_y \sp{B}_z \\
    -i\omega \sp{E}_{xz} &= -c^2\partial_z \sp{B}_y - \frac{\sigma(z)}{\epsilon_0}\sp{E}_{xz}\\
    -i\omega \sp{E}_{yz} &= c^2\partial_z \sp{B}_x - \frac{\sigma(z)}{\epsilon_0}\sp{E}_{yz}\\
    -i\omega \sp{E}_{yx} &= -c^2ik_x \sp{B}_z \\
    -i\omega \sp{E}_z &= c^2(ik_x \sp{B}_y - ik_y \sp{B}_x) - \frac{\alpha(z)q}{\epsilon_0} e^{i\omega z/v}
    \label{eq:FourierMaxwellEz}
    \end{align}
\end{subequations}
\begin{subequations}
    \label{eq:FourierMaxwellB}
    \begin{align}
    -i\omega \sp{B}_{xy} &= -ik_y \sp{E}_z \\
    -i\omega \sp{B}_{xz} &= \partial_z \sp{E}_y - \frac{\sigma(z)}{\epsilon_0} \sp{B}_{xz} \\
    -i\omega \sp{B}_{yz} &= -\partial_z \sp{E}_x - \frac{\sigma(z)}{\epsilon_0} \sp{B}_{yz} \\
    -i\omega \sp{B}_{yx} &= ik_x \sp{E}_z \\
    -i\omega \sp{B}_z &= -(ik_x \sp{E}_y - ik_y \sp{E}_x)
    \label{eq:FourierMaxwellBz}
    \end{align}
\end{subequations}
\begin{subequations}
    \label{eq:FourierMaxwellSum}
    \begin{align}
    \sp{E}_{x} &= \sp{E}_{xy} + \sp{E}_{xz} \\
    \sp{E}_{y} &= \sp{E}_{yz} + \sp{E}_{yx} \\
    \sp{B}_{x} &= \sp{B}_{xy} + \sp{B}_{xz} \\
    \sp{B}_{y} &= \sp{B}_{yz} + \sp{B}_{yx}
    \end{align}
\end{subequations}

The Fourier transform of the Ansatz fields in the PML \cref{eq:PMLfield} is
(see \cref{sec:Fourier} for a demonstration)
\begin{subequations}
\label{eq:FourierPMLfield}
\begin{align}
\vecthree{\sp{E}_x}{\sp{E}_y}{\sp{E}_z} = -i\alpha(z)\frac{q}{\epsilon_0 v}
\frac{e^{i\omega z/v}}{[k_x^2 + k_y^2 + \omega^2/(\gamma v)^2]}
\vecthree{k_x}{k_y}{\omega/(\gamma^2 v)} \\
\vecthree{\sp{B}_x}{\sp{B}_y}{\sp{B}_z} = -i\alpha(z)\frac{q}{\epsilon_0 c^2}
\frac{e^{i\omega z/v}}{[k_x^2 + k_y^2 + \omega^2/(\gamma v)^2]}
\vecthree{-k_y}{k_x}{0}
\end{align}
\end{subequations}

\subsection{Condition on $\alpha(z)$}
\label{sec:alphaCondition}

From the above expressions, it can be seen that the Ansatz fields in
Fourier space \cref{eq:FourierPMLfield} automatically satisfy
the longitudinal component of the PML equation
\cref{eq:FourierMaxwellEz,eq:FourierMaxwellBz}. In addition, inserting \cref{eq:FourierPMLfield}
into the other equations of \cref{eq:FourierMaxwellE,eq:FourierMaxwellB}
provides the expression of the split components of $\vec{E}$ and $\vec{B}$:
\begin{equation}
\label{eq:FourierSplitE}
\vecfour{\sp{E}_{xy}}{\sp{E}_{xz}}{\sp{E}_{yz}}{\sp{E}_{yx}} =
- i \alpha(z)\frac{q}{\epsilon_0 v}
\frac{e^{i\omega z/v}}{[k_x^2 + k_y^2 + \omega^2/(\gamma v)^2]}
\vecfour{0}{\chi(z)k_x}{\chi(z)k_y}{0}
\end{equation}
\begin{equation}
\label{eq:FourierSplitB}
\vecfour{\sp{B}_{xy}}{\sp{B}_{xz}}{\sp{B}_{yz}}{\sp{B}_{yx}} =
-i\alpha(z)\frac{q}{\epsilon_0 v^2}
\frac{e^{i\omega z/v}}{[k_x^2 + k_y^2 + \omega^2/(\gamma v)^2]}
\vecfour{k_y/\gamma^2}{-\chi(z)k_y}{\chi(z)k_x}{-k_x/\gamma^2}
\end{equation}
where:
\begin{equation}
\chi(z) \equiv \frac{i\omega + \frac{v}{\alpha}\frac{d \alpha}{dz}}{i\omega - \frac{\sigma}{\epsilon_0}}
\end{equation}
In order to show that the Ansatz fields \cref{eq:FourierPMLfield} satisfy
the PML equations, it now suffices to show that
\cref{eq:FourierPMLfield,eq:FourierSplitE,eq:FourierSplitB} satisfy the remaining
set of PML equations \cref{eq:FourierMaxwellSum}. It is easy to show that
\cref{eq:FourierMaxwellSum} is indeed satisfied if and only if $\chi(z)=1$, i.e.
if and only if:
\begin{equation}
\frac{d\alpha(z)}{dz} = -\frac{\sigma(z)}{\epsilon_0 v}\alpha(z)
\end{equation}
The solution to this differential equation, with the continuity condition
$\alpha(0)=1$ (see \cref{eq:continuitycondition}), is indeed the expression given in
\cref{eq:alpha}.

\section{Fourier transform of the fields associated with a
relativistic particle}
\label{sec:Fourier}

It is well-known that, in 3 dimensions, the Fourier transform of a
$1/r$ field has the following form (see e.g. \cite{appel2007})
\begin{equation}
\int_{\mathbb{R}^3} d\vec{X} \;\frac{1}{|\vec{X}|}e^{-i \vec{K}\cdot \vec{X}} = \frac{4\pi}{\vec{K}^2}
\end{equation}
Multiplying on both sides by $-i\vec{K}$ and using integration by parts:
\begin{equation}
-\int_{\mathbb{R}^3} d\vec{X} \;
\left(\frac{\partial \;}{\partial \vec{X}}\frac{1}{|\vec{X}|}\right)e^{-i \vec{K}\cdot \vec{X}} =
-i \frac{4\pi\vec{K}}{\vec{K}^2}
\end{equation}
and then upon carrying out the derivative, and using the vector notation $\vec{X}=\vecthree{X_1}{X_2}{X_3}$
and $\vec{K}=\vecthree{K_1}{K_2}{K_3}$:
\begin{align}
\iiint dX_1\,dX_2\,dX_3\;&
\frac{e^{-i (K_1 X_1 + K_2 X_2 + K_3 X_3)}}{(X_1^2 + X_2^2 + X_3^2)^{3/2}}\vecthree{X_1}{X_2}{X_3}\nonumber\\
&= -i \frac{4\pi}{(K_1^2 + K_2^2 + K_3^2)}\vecthree{K_1}{K_2}{K_3}
\end{align}
We now use the following change of variable (at fixed $z$):
\begin{align}
&X_1 = x\qquad &X_2 = y \qquad &X_3 = \gamma(z-vt) \\
&K_1 = k_x \qquad &K_2 = k_y \qquad &K_3 = \omega/(\gamma v)
\end{align}
and obtain:
\begin{align}
\gamma v \iiint dx\,dy\,dt\;&
\frac{e^{-i [k_x x + k_y y + \omega(z/v - t)]}}{[x^2 + y^2 + \gamma^2(z-vt)^2]^{3/2}}\vecthree{x}{y}{\gamma(z-vt)}\nonumber\\
&= -i \frac{4\pi}{[k_x^2 + k_y^2 + \omega^2/(\gamma v)^2]}\vecthree{k_x}{k_y}{\omega/(\gamma v)}
\end{align}
Finally, by multiplying the third vector component of this equation by $1/\gamma$, and multiplying
all components by various physical factors:
\begin{align}
\iiint dx\,dy\,dt&\;\frac{\gamma q}{4\pi \epsilon_0}
\frac{e^{i\omega t - ik_x x - ik_y y}}{[x^2 + y^2 + \gamma^2(z-vt)^2]^{3/2}}\vecthree{x}{y}{z-vt}\nonumber\\
&= -i \frac{q}{\epsilon_0 v}\frac{e^{i\omega z/v}}{[k_x^2 + k_y^2 + \omega^2/(\gamma v)^2]}\vecthree{k_x}{k_y}{\omega/(\gamma^2 v)}
\end{align}

\bibliography{Bibliography.bib}

\end{document}